\documentclass[prl,twocolumn,showpacs,preprintnumbers,amsmath,amssymb]{revtex4}
\usepackage{graphicx}
\begin{document}

\title{Correlated magnetic states in domain and grain boundaries in graphene}

\author{Simone S. Alexandre$^1$, A. D. L\'ucio$^2$, A. H. Castro Neto$^{3,\ast}$, and R. W. Nunes$^1$}
\affiliation{$^1$Departamento de F\'\i sica, ICEx, Universidade
Federal de Minas Gerais, 31270-901, Belo Horizonte, MG, Brazil\\
$^2$Departamento de Ci\^encias Exatas, Universidade
Federal de Lavras, 37200-000, Lavras, MG, Brazil\\
$^3$Graphene Research Centre and Department of Physics
National University of Singapore, 2 Science Drive 3, Singapore, 117542}
\date{\today}

\begin{abstract} 
{\it Ab initio} calculations indicate that while the electronic states
introduced by grain boundaries in graphene are only partially confined
to the defect core, a domain boundary introduces states near the Fermi
level that are very strongly confined to the core of the defect, and
that display a ferromagnetic ground state. The domain boundary is
fully immersed within the graphene matrix, hence this magnetic state
is protected from reconstruction effects that have hampered
experimental detection in the case of ribbon edge states. Furthermore,
our calculations suggest that charge transfer between one-dimensional
extended defects and the bulk in graphene is short ranged for both
grain and domain boundaries.
\end{abstract}

\pacs {73.22.-f, 73.20.Hb, 71.55.-i}

\maketitle

\begin{figure}
\includegraphics[width=8.0cm]{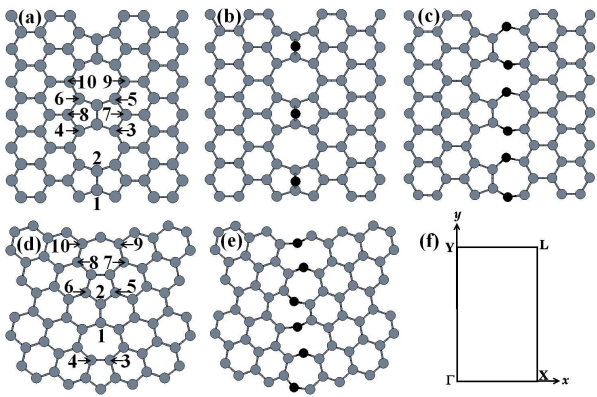}
\caption{Geometries of extended one-dimensional defects in
graphene. Carbon (oxygen) atoms are shown as grey (black) circles. (a)
DB558: a domain boundary. (b) DB558+O: oxidized form with one O atom per
defect unit. (c) DB558+2O: oxidized form with two O atoms per defect
unit. (d) GB: a large-angle tilt grain boundary. (e) GB+O: oxidized
form of the GB. (f) Schematic view of the first quadrant of the
Brillouin zone associated to the defect supercells. High-symmetry
points and Cartesian axes are indicated.}
\label{fig1}
\end{figure}

Controlling electronic transport and tailoring magnetic states at the
nanoscale in graphene rank among the main issues related to the
prospective application of this material in nanoelectronics and
spintronics. In particular, electronic and magnetic states of extended
line defects in graphene have been considered as possible conducting
one-dimensional (1D) electronic channels and platforms for tailored
spin states for spintronic applications.~\cite{ahcn1,lahiri} Two
recent developments highlight the focus on extended line defects: (i)
the recognition that mass-scale production of graphene should
inevitably lead to a polycrystalline material, containing
one-dimensional grain boundaries (GB)~\cite{yu}; and (ii) the recent
theoretical prediction~\cite{umesh} and experimental realization of
domain boundaries (DB) in graphene by controlled deposition of the
material on metallic substrates~\cite{lahiri}. In the case of GB's, a
large volume of works has accumulated in the last few
years~\cite{jsa,simonis,cervenka,louie1}, while for the DB
produced in Lahiri {\it et al.} experiment, valley-filter
properties~\cite{gunlycke} and its effects on the magnetic edge states
of a graphene ribbon~\cite{lin} have been theoretically
investigated. Further, when chemical reduction of a readily available
material, such as graphene oxide, is used as a viable route for
mass-scale graphene synthesis~\cite{cristina,bagri}, the presence of
residual functional groups should affect the material electronic and
magnetic properties.

In this context, the nature of the electronic states introduced by
such extended 1D defects in graphene is a topic that deserves close
inspection. More specifically, whether GB's and DB's act as quasi-1D
conducting channels immersed in the bulk of graphene is the question
we seek to address in this work. We also consider the issue of
self-doping in graphene induced by the presence of such extended 1D
defects~\cite{prb06}, that occurs when the line defect attracts charge
carriers, resulting in charged defective lines surrounded by a doped
graphene matrix.

Our calculations indicate that while the electronic states introduced
by a model GB structure in graphene hybridize with the bulk states and
are only partially confined to the defect core, the DB defect
introduces a sharp resonance in the density of states (DOS) of
graphene, that lies just above the Fermi level in the neutral system,
and is associated to electronic states that are very strongly
confined to the core of the defect. Our results suggest that, when a
graphene sample containing a DB is doped and these quasi-1D states are
populated, a ferromagnetic state is realized, which is confined to
zigzag chains along the defect core that are fully immersed within a
bulk graphene matrix. The quantum confinement induced by the presence
of a domain boundary leads to an enhancement of the Coulomb
interactions and stronger electron-electron correlations. Because of
their 1D nature, these correlated states do not show long range
order. Instead, they present power law or algebraic correlation
functions. Furthermore, given that these 1D states are immersed in a
metallic environment, they are unique examples of {\it open Luttinger
liquids}~\cite{oll}. 

Moreover, we find that charge transfer between bulk graphene and the
line defects is essentially local, with charge redistribution taking
place within a region of $\sim$3-5~\AA\ from the geometric center of
the 1D defects. Both this charge-transfer region and the surrounding
bulk regions remain essentially neutral. The absence of a charge
monopole moment means that these 1D defects should act as weak charge
carrier scatterers, in agreement with recent transport experiments in
graphene grown by chemical vapor deposition (CVD)~\cite{cvd}.

In this study, we examine the nature of the electronic states
introduced by the two aforementioned types of extended 1D defects in
graphene: (i) large-angle tilt grain
boundaries~\cite{simonis,cervenka}, expected to occur commonly in
mass-scale production of graphene; (ii) domain boundaries produced by
deposition of a graphene layer on a Nickel
substrate~\cite{umesh,lahiri}. For both 1D defects, we also consider
oxidized forms, with oxygen atoms bound to the atoms at the core of
the defect. Figure~\ref{fig1}(a) shows the atomistic model for the
domain boundary defect, which consists of periodic units composed of
one octagon and two side-sharing pentagons along the defect line,
which we label DB558 in the following discussion. The oxidized forms
with one (DB558+O) and two oxygen atoms (DB558+2O) per defect unit are
shown respectively in Figs.~\ref{fig1}(b) and
(c). Figure~\ref{fig1}(d) shows the GB model proposed in
Ref.~\onlinecite{simonis}, while Fig.~\ref{fig1}(e) shows the fully
unzipped oxidized form of this defect (GB+O).

Our calculations are performed in the framework of Kohn-Sham density
functional theory (DFT), within the generalized-gradient approximation
(GGA)~\cite{ks-gga} and norm-conserving pseudopotentials in the
Kleinman-Bylander factorized form~\cite{tm-kb}. We use the LCAO method
implemented in the SIESTA code~\cite{siesta}, with a double-zeta basis
set plus polarization orbitals. A Mulliken charge partition is
employed for the analysis of the charge-transfer between bulk and
defects. Supercells of 66 (84) atoms are employed for the DB558 (GB)
calculations.  We performed convergence tests to ensure that, at these
cell sizes, our results are not affected by interaction between the
defects and their periodic images.

\begin{figure}
\includegraphics[width=8.5cm]{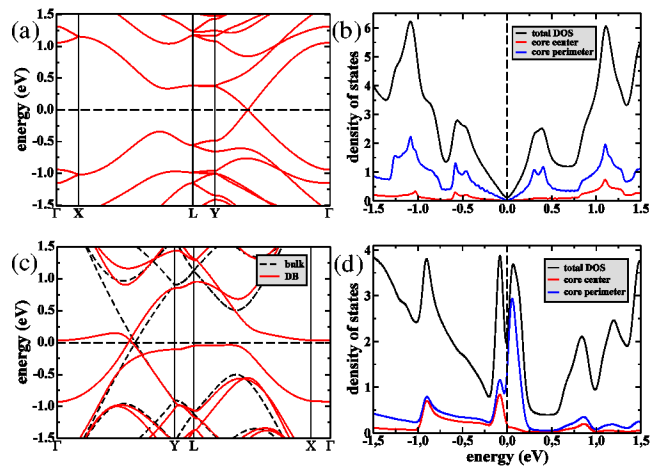}
\caption{(Color online) (a) Band-structure for the GB along the
Brillouin-zone lines in Fig.~1(f). (b) DOS for the GB. (c)
Band-structure for the DB558. (d) DOS for the DB558.}
\label{fig2}
\end{figure}

In the case of GB's in graphene, the electronic structure has been
amply discussed in Ref.~\onlinecite{jsa}, where the occurrence of an
anisotropic Dirac cone in a Brillouin-zone (BZ) line along the GB
direction was analyzed. In Ref.~\onlinecite{jsa}, the GB electronic
states near the Fermi level (FL) were found to disperse in all
directions, indicating hybridization with bulk states. Further
indication of such hybridized nature of GB states is provided by the
analysis of the full band structure and density of states (DOS) of the
GB supercell. Figure~\ref{fig2}(a) shows the band structure along the
BZ lines shown in Fig.~\ref{fig1}(f). Note that the states near the
Dirac point show sizeable dispersions in directions perpendicular to
the defect (Y-M and $\Gamma$-X lines). The contributions of the
orbitals centered on the atoms at the GB core to the resonances
introduced in the DOS of graphene are shown in
Figure~\ref{fig2}(b). For the purpose of our analysis, we define the
core of both the GB and DB558 defects as composed by the ten atoms
that form the topological defects along the defect line, as indicated
in Fig.~\ref{fig1}. In the DOS plots, we refer to atoms 1 and 2, that
are placed along the line at the geometric centers of both defects, as
core center, while the remaining eight core atoms are referred to as
core perimeter. In the case of the GB, the contribution of the ten
core atoms to the four DOS peaks near the FL ranges from 41\% to
57\%. These results will be contrasted with the case of the DB558 in
the following.

The band structure and the DOS for the DB558 are shown in
Fig.~\ref{fig2}(c) and (d), respectively. In Fig.~\ref{fig2}(c) we
also include the band structure of a 64-atom bulk supercell, obtained
by removing core atoms 1 and 2 from the geometry in
Fig.~\ref{fig1}(a). Given the periodicity of both defect and bulk
cells along the DB558 direction (a zigzag direction of the bulk
matrix), the Dirac point of the bulk cell folds onto the point at
$2\pi/3\ell$, where $\ell$ is the period of the
DB558~\cite{nakada}. Note that, to a large degree, the changes in the
DB558 band structure with respect to the bulk one are concentrated in
a region of $\pm 0.5 eV$ from the FL, and that within $\sim$0.2~eV
above (below) the FL we find bands with flat sections along the
$\Gamma-$Y (L-X) direction (both parallel to the defect line). These
states show little or no dispersion along the $\Gamma$-X and Y-M lines
that are perpendicular to the defect line. The flat-band character of
the states above the FL, near the zone center, leads to a
ferromagnetic state for a bulk 1D defect in graphene consisting
entirely of carbon atoms, as discussed below.

Considering now the DOS for the DB558 in Fig.~\ref{fig2}(d), we find
two sharp resonances within $\sim\pm$0.1~eV from the FL, associated to
the 1D defect. These resonances reflect the presence of extended van
Hove regions in the in the bandstructure, as seen in Fig. 2(c). We
observe that in one feature the DOS of the DB558 differs markedly from
the GB one: the peak just above the Fermi level shows a very strong
concentration on the core atoms, with $\sim$80\% of the total DOS
concentrated on the orbitals centered on the perimeter core atoms
numbered 3, 4, 5, and 6, along the zigzag chains that bond to the
dimers (atoms 1 and 2) at the core center in
Fig.~\ref{fig1}(b). Adding the contributions of the atoms from the
same sublattice as atoms 3-6, in the zigzag chains nearest to the core
on both sides of the defect line, we already account for $\sim$96\% of
the total DOS for this peak. A strongly one-dimensionally confined
empty state along the defect core is thus a characteristic of the
DB558. In our calculations, however, these states are empty in the
neutral system, hence no spin polarization is induced.
\begin{figure}
\includegraphics[width=7.5cm]{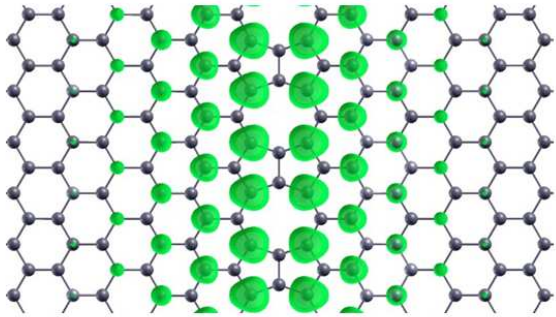}
\caption{(Color online) Isosurface of spin polarization density for
domain-boundary magnetic state.}
\label{fig3}
\end{figure}

In order to investigate the formation of magnetic states in gate-doped
versions of this system, we added one electron per supercell, a doping
concentration that raises the FL by about $\sim$0.07~eV. A
spin-polarized calculation stabilizes a ferromagnetic state, with a
magnetic moment of 0.52 $\mu_B$ per defect unit, and a formation
energy that is $\sim$40 meV per defect unit lower in energy than the
unpolarized state. Figure~\ref{fig3} shows a representative isosurface
of the difference between majority and minority spin densities. The
spin density is concentrated on the sublattice of the atoms 3-6, with
a negligible contribution from the dimer atoms themselves and also
from the atoms near the core that belong to the other sublattice. We
see here a manifestation of a magnetic state along a line defect that
is fully immersed within the bulk of graphene. The fact that these
states have a negligible contribution along the dimer atoms is perhaps
the origin of the topological disruption in the electronic states that
leads to the localization of the related $\pi$-orbital bands and hence
to the magnetic state. We speculate that such DB magnetic states
should be more easily detectable experimentally than those predicted
to occur along the edges in graphene
ribbons~\cite{nakada,hlee,louie3,simone}, since the zigzag chains
along the DB are protected from reconstruction.
\begin{figure}
\includegraphics[width=8.5cm]{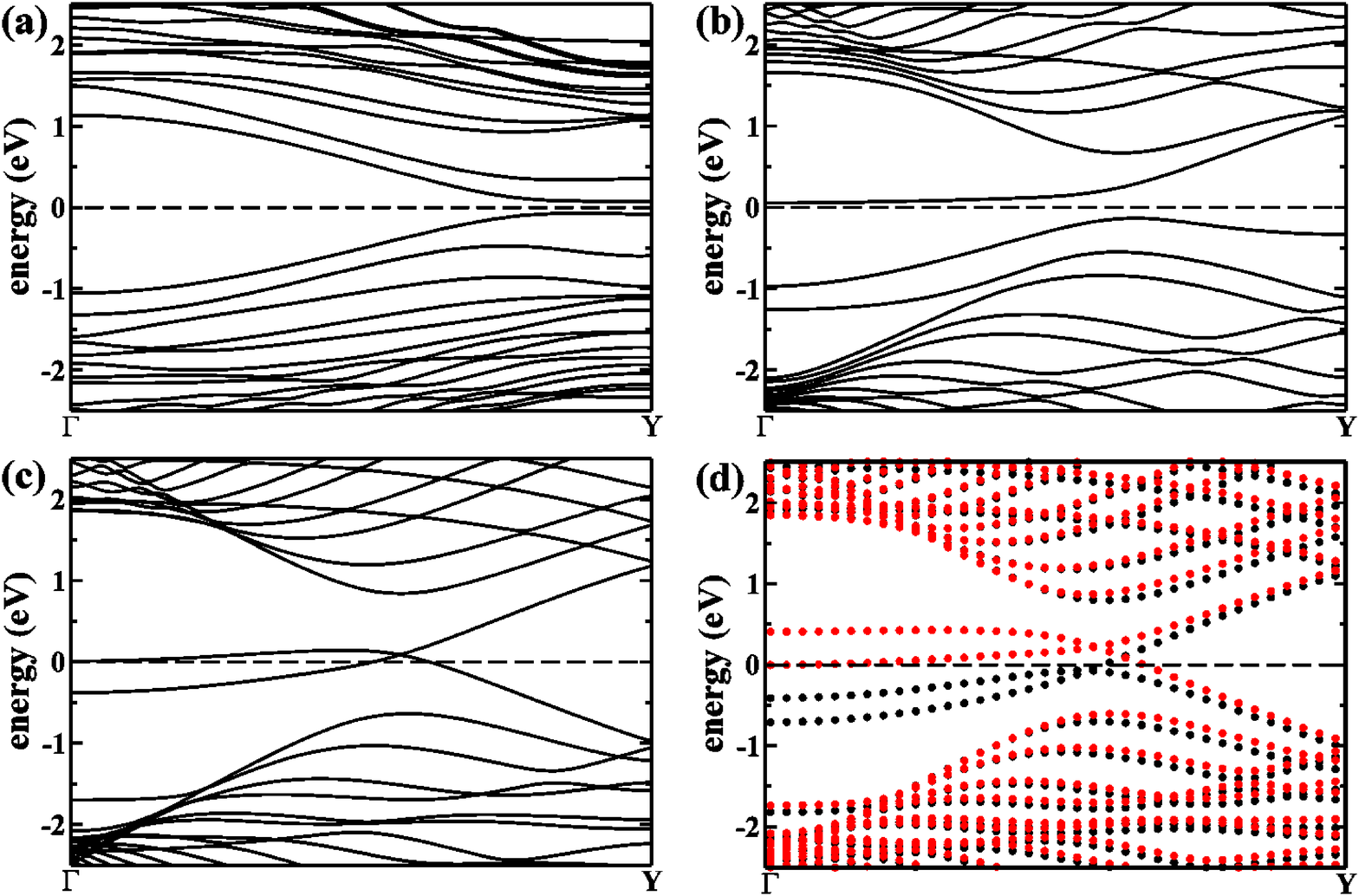}
\caption{(Color online) (a) Band-structure for the GB+O. (b)
Band-structure for the DB558+2O. (c) Band-structure for the
DB558+O. (d) Band-structure for the DB558+O, including spin
polarization. Black (red) symbols show majority (minority) spin
bands.} 
\label{fig4}
\end{figure}

To address the effect of chemical doping on the electronic states of
the 1D defects, we considered the oxidized forms shown in
Fig.~\ref{fig1}. The corresponding band structures are shown in
Fig.~\ref{fig4}. For both the GB+O and DB55+2O, where oxygen-induced
unzipping of the C-C bonds takes place, we observe the opening of
small gaps of 55 meV and 190 meV, respectively. In neither of these
two cases we find magnetic ground states in the neutral system. An
interesting case is the DB558+O system, with one oxygen atom per
defect cell, in which the oxygen atom bonds with the two carbon atoms
at the defect core in the bridge position. In this case, as shown in
Fig.~\ref{fig4}(c), the occupied band along the $\Gamma -$Y line that
is $\sim$0.1~eV below the FL in the pure DB558 system is now deeper in
energy. More importantly, a significant change is observed in the flat
band of confined 1D states near $\Gamma$ that now crosses the FL.  As
a result, a magnetic ground state is stabilized, with a net magnetic
moment of 1.1$\mu_B$ per unit cell, as shown in Fig.~\ref{fig4}(d),
where majority and minority spin bands are plotted. Note that
spin-polarization is strongly localized on the 1D defect, with
significant exchange splitting restricted to the quasi-1D bands
crossing the FL. This ferromagnetic ground-state is 43 meV (per defect
unit) lower in energy than the non-polarized state. This value is very
close to that for the pure DB558 system with an extra electron per
cell. Hence, chemical doping offers another way of stabilizing this
ferromagnetic state. The ferromagnetic instability of these flat-band
states is further confirmed by a calculation we performed for the
DB558+2O system with an extra charge added to the defect
supercell. Again, the FL shifts into the band of quasi-1D states, and
the system stabilizes a ferromagnetic state. Just as in the case
without O doping, the presence of extended van Hove regions is also
clearly seen in the band structures in Fig.~\ref{fig4}.

\begin{figure}
\includegraphics[width=9.0cm]{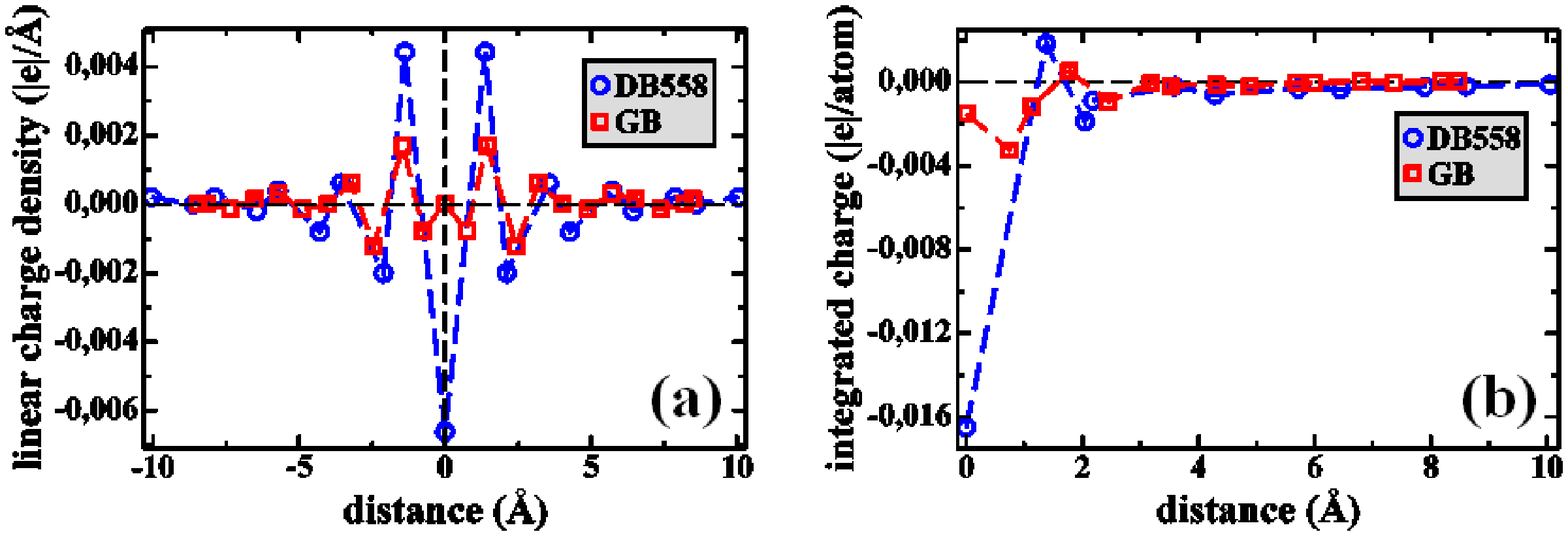}
\caption{(Color online) (a) Charge density distribution for the DB558
(blue circles) and the GB (red squares). (b) Integrated charge $I(r)$
(see text) for the DB558 and the GB.}
\label{fig5}
\end{figure}

We turn our attention now to the charge transfer between the 1D
defects and the surrounding graphene bulk, since this may give
important qualitative information about the nature of the scattering
of charge carriers by such 1D defects. In Fig.~\ref{fig5}, we show the
profile of the charge distribution around the 1D defects, and the
integrated charge per atom, both as a function of the distance to the
geometric center of the defect, for the DB558 and the GB. The charge
distribution is shown as linear charge densities representing the net
charge summed over atoms at the same distance from the 1D defect,
divided by the period of the defect unit. The integrated charge is the
integral of this linear charge density profile, computed as total net
charge per atom, and defined as follows:

\begin{eqnarray}
I\left(R\right) = \frac{1}{N(R)}\sum_{r_i = 0}^R
\lambda\left(r_i\right)\times \ell_{d}\;;
\end{eqnarray}
where $\lambda\left(r_i\right)$ is the linear charge density for the
line of atoms at a distance $r_i$ from the defect,
$\ell_{d}$ is the period of the 1D defect, and $N\left(R\right)$ is the
number of atoms summed over all lines with $r_i\le R$.

Figure 5 shows that the charge redistribution around the 1D defects is
non-monotonic, with the linear charge distributions alternating in
sign on both sides of the defect, as seen in Fig.~\ref{fig5}(a). Note
also that charge transfer between the 1D defects and the graphene
matrix occurs in a range of $\sim$3-5~\AA\ from the defect line, with
the graphene matrix becoming neutral beyond this range, as shown in
the plots for $I(R)$ in Fig.~\ref{fig5}(b).  Since the charge
distribution region recovers neutrality within $\sim$3-5~\AA\ from the
defect center, we expect no long range Coulomb scattering of charge
carriers from both 1D defects. In our calculations, oxidation of the
1D defects is found not to affect this localized character of the
charge balance between the defects and the graphene bulk. Such
extended 1D defects should then play a minor role as a source of
carrier scattering in graphene.

In summary, we find that the electronic states introduced by grain
boundaries in graphene are only partially confined to the defect core,
while a domain boundary introduces unoccupied electronic states near
the Fermi level that are very strongly confined to the core of the
defect, and that, when populated by doping, display a ferromagnetic
ground state in a 1D defect that is fully contained within the bulk
matrix and that consists entirely of carbon atoms. Being fully
bulk-immersed, this ferromagnetic state is protected from
reconstruction and should be more easily detectable experimentally
than those predicted to exist along the edges of graphene
ribbons. Furthermore, our calculations indicate that charge transfer
between bulk graphene and both types of extended 1D defects is
strongly localized, with charge redistribution confined to
$\sim$3-5~\AA\ from the geometric center of the 1D defect, implying
that these 1D defects should act as weak charge scatterers in
graphene.

$\ast$ On leave from Boston University, USA.

\begin{acknowledgments}
SSA, RWN, and ALD acknowledge support from Brazilian agencies CNPq,
FAPEMIG, and Instituto do Mil\^enio em Nanoci\^encias-MCT. AHCN thanks
the financial support of the NRF-CRP award ¡ÈNovel 2D Materials with
Tailored Properties: Beyond Graphene¡É (R-144-000-295-282), US/DOE
grant DE-FG02-08ER46512, and US/ONR grant MURI N00014-09-1-1063.
\end{acknowledgments}

\end{document}